\documentclass[10pt]{revtex4}
\usepackage{hyperref}
 
\begin{document}

\title{\Large Emergence of Lowenstein-Zimmermann mass terms for QED$_{3}$}

 \author{ Sudhaker Upadhyay\footnote {e-mail address: sudhakerupadhyay@gmail.com}}
\author{ Manoj Kumar Dwivedi\footnote {e-mail address: manojdwivedi84@gmail.com}}
\author{ Bhabani Prasad Mandal\footnote {e-mail address: bhabani.mandal@gmail.com}}

\affiliation { Department of Physics, 
Banaras Hindu University, 
Varanasi-221005, INDIA.  }

\begin{abstract}
 In this paper we consider 
 a super-renormalizable theory of massless QED in $(2+1)$ dimensions and discuss their BRST symmetry transformation.  By extending  the BRST transformation we derive the Nielsen
 identities for the theory. Further, we
 compute the generalized BRST (so-called FFBRST) transformation by making the
 transformation parameter field-dependent. Remarkably, we observe that the 
 Lowenstein-Zimmerman mass terms,
 containing Lowenstein-Zimmerman parameter which plays an important role in the BPHZL
 renormalization program, along with the external sources coupled to the non-linear BRST 
 variations appear  naturally in the theory through a  FFBRST  transformation. 
\end{abstract}
 
  \maketitle  
  
\textit{Keywords}: QED$_{3}$;  Lowenstein-Zimmermann mass; BRST symmetry. 

\textit{PACS}: 11.15.-q; 11.15.Wx.
\section{Introduction}
Massless QED in $2+1$ dimensions (QED$_3$) has very interesting and crucial features
 in the frontier research. Recently in a seminal work \cite{del}, QED$_3$
 has been established a super-renormalizable theory utilizing a powerful algebraic renormalization method.
Also, it is a ultraviolet finite and  parity invariant  theory \cite{gua,gua1,gua2,gua3,del}. The massless QED$_3$ provides
an ideal platform to tackle the infrared divergence present in the theory.
  The parity anomaly 
 has been dismissed for such theories at all orders of perturbation.
The  Lowenstein-Zimmerman scheme plays an important role in algebraic proof of ultraviolet and infrared finiteness and in the dismissal of parity anomaly. In spite of that the QED$_3$ also gets relevance in the study of high-$T_c$ superconductivity \cite{nd,kf}.
The dynamical fermion mass generation and chiral symmetry breaking 
for QED$_3$ are studied in \cite{pen}. In this context, the dynamical   mass generation 
using Hamiltonian lattice methods has  also been investigated which has been found 
in agreement with both the strong coupling expansion and with the Euclidean
lattice simulations \cite{lee}.

The  Lowenstein-Zimmermann subtraction scheme plays a major role in the algebraic proof  on the ultraviolet and infrared finiteness, and to show the absence
of a parity and infrared anomaly, in the massless QED$_3$,
which is based on general theorems of perturbative quantum field
theory \cite{jh,jh1,jh2,jh3,jh4,brs,brs1,brs2,jh5}.
The  massless QED$_3$ with and without 
   Lowenstein-Zimmermann mass terms is a gauge theory and in order to deal   such theories
   we need to break the gauge invariance by fixing the gauge.
The gauge invariance of the quantum
action is also essential because the expectation values of physical quantities become
independent of the choice of the gauge-fixing term. To realize the gauge independence i.e. the estimating S-matrix elements (or expectation values) of 
the gauge independent quantities, one utilizes the  on-shell quantum effective action (i.e., evaluated at those configurations that extremize it). 
 Nielsen identities \cite{nil} suggest that the variation of the quantum effective
action due to changes in the functions that fix the gauge must be linear in the quantum corrected
equations of motion for the mean fields. This follows that the on-shell quantum
effective action does not depend on the choice of the gauge breaking term. Although the mean fields  depend on the gauge-fixing,  this dependence gets canceled by the explicit gauge-fixing dependence of the quantum effective action \cite{del1,del2,del3}.
 
In the proof of algebraic renormalizability  the BRST symmetry has an incredible importance
\cite{masud}.
 The generalizations of BRST symmetry have been studied in various contexts
  \cite{bm,mru,sdj1,rb,susk,jog,sb1,sb2,sb3,sb4,sb5,sb6,sb7,sb8,smm,sudd,fsm,fsm1,sud,sud1,sud2,sud3,
  sud4,sud5,sud6,sud7, fs,rbs,rbs1} since their introduction in 
  Ref. \cite{bm}. 
 For examples,  a correct prescription for   poles in the gauge field propagators in noncovariant gauges has been derived by connecting  covariant gauges and noncovariant gauges of the theory by using FFBRST transformation \cite{jog}.
 The long outstanding problem of divergent energy integrals in Coulomb gauge has been
 regularized using FFBRST transformation \cite{sdj1}.
The Gribov-Zwanziger theory \cite{gri,zwan},  a limiting case of Yang-Mills theory ,
 which  plays a crucial role in the non-perturbative
low-energy region while it can be neglected in the  perturbative high-energy
region, has also been related to the YM theory  in Euclidean space through FFBRST transformation \cite{sb, sb20}. 
The FFBRST formulation has also been established at quantum level
utilizing BV formulation \cite{mru}. 
Recently, the field-dependent BRST transformation has also been considered with same philosophy and goal as in original work \cite{bm} even though in slightly
different manner to calculate the explicit Jacobian for such transformation \cite{lav,ale,ale1,ale2}.   A systematic Hamiltonian formulation of such theories have also been done \cite{bata,bata1}.

 In this paper we analyse the  massless QED$_3$, which is a super-renormalizable and free from 
 parity and infrared anomaly, within generalized  BRST (FFBRST) framework.
 To implement the FFBRST formulation, we first make the infinitesimal parameter of transformation field dependent through continuous interpolation of a
 parameter $\kappa: (0\leq \kappa\leq 1)$. Then we integrate the infinitesimal field-dependent transformation parameter with respect to $\kappa$ in its extreme limit
 to obtain the FFBRST transformation. 
 Such FFBRST transformation leads a non-trivial Jacobian for the path integral measure
 of functional integral. 
 Remarkably, we realize that the Lowenstein-Zimmerman mass terms along with the external sources for 
 QED$_3$ emerges naturally through Jacobian calculation
 under FFBRST transformation.   In proof of renormalizability of the theory such source term 
play an important role. 
 
The plan of the paper is as follows. We start with a brief discussion of massless QED$_3$ theory in Sec. II.
The section III is devoted to derive the Nielsen identities.
Further, in section IV,  we sketch the generalized BRST transformation. 
In Sec. V, we show the emergence of  Lowenstein-Zimmermann mass terms along with
external source term for  QED$_{3}$.   The last section is reserved for conclusions. 
  \section{The massless QED$_{3}$}
  In this section, we recapitulate the massless QED$_3$ with and without 
   Lowenstein-Zimmermann mass terms. 
Let us begin with the gauge invariant massless action for QED$_3$ defined as 
\begin{eqnarray}
\Sigma_{0}=\int  d^3x\left[-\frac{1}{4} F_{\mu\nu}F^{\mu\nu} + i\bar\psi  \gamma^\mu D_\mu\psi   \right],
\end{eqnarray}
where the covariant derivative is defined by $D_\mu=\partial_\mu +ieA_\mu$ and $e$ is a dimensionful 
coupling constant.
Since, the gauge invariant theory can be quantized correctly only after
choosing a particular gauge for the theory. We define
the gauge-fixing and induce ghost terms for QED$_3$  as follows:
\begin{equation}
\Sigma_{gf+gh} =\int  d^3x\left[  b\partial^\mu A_\mu + \frac{\xi}{2}b^2 + \bar c \partial_\mu 
\partial^\mu c \right],
\end{equation}
where $b, c$ and $\bar c$ are Nakanishi-Lautrup auxiliary field, ghost field and anti-ghost field respectively.
Therefore, the  effective action can be written easily by
\begin{eqnarray}
\Sigma_{eff}=\Sigma_{0}+\Sigma_{gf+gh}.
\end{eqnarray}
This quantum action remains invariant under following nilpotent BRST  transformations:
\begin{eqnarray}
\delta_b A_\mu &=&s_b A_\mu \ \epsilon = \frac{1}{e} \partial_\mu c\ \epsilon, \nonumber\\ 
\delta_b \psi  &=&s_b \psi \ \epsilon =  ic\psi\ \epsilon, \nonumber\\  
\delta_b \bar\psi  &=&s_b \bar{\psi} \ \epsilon =  -ic\bar\psi\ \epsilon, \nonumber\\  
\delta_b c  &=& s_b c \ \epsilon = 0, \nonumber\\  
\delta_b \bar c  &=&s_b \bar c \ \epsilon =  \frac{1}{e} b\  \epsilon, \nonumber\\  
\delta_b b  &=&s_b b \ \epsilon =  0, \label{BRS}
\end{eqnarray}
where $\epsilon$ is an infinitesimal anticommuting parameter of transformation.

Now, we introduce a gauge invariant  Lowenstein-Zimmermann mass term  for
the massless QED$_3$ which makes the theory super-renormalizable  
 \cite{del}:
\begin{eqnarray}
\Sigma_m = \int  d^3x\left[  \frac{\mu}{2}(s-1)\epsilon^{\mu\nu\rho}A_\mu \partial_\nu A_\rho + {m}(s-1)\bar\psi \psi \right] ,   
\end{eqnarray}
where $0\leq s\leq 1$ is the  Lowenstein-Zimmermann parameter which plays a central role 
in BPHZL renormalization program.
Now, the effective action for  QED$_3$ with the Lowenstein-Zimmermann mass terms is given by 
\begin{equation}
\Sigma_{eff}^m=\Sigma_{eff}+\Sigma_m.
\end{equation}
It is desirable to define an external source term  for the theory  \cite{del},
\begin{eqnarray}
\Sigma_{ext}= \int d^3x [ \bar{\Omega}s_b\psi - s_b\bar{\psi}\Omega].
\end{eqnarray}
This source terms play  an important role in demonstrating the Slavnov-Taylor identity
which guarantees the renormlizability of the theory. With this source term 
the final action reads,
\begin{eqnarray}
\Sigma_T=\Sigma_{eff}+\Sigma_m +\Sigma_{ext}.\label{eff}
\end{eqnarray}
This final action ($\Sigma_T$) is invariant under the same set of BRST
transformations (\ref{BRS}). Now we would like to derive the Nielsen identities
for the  QED$_3$ with the Lowenstein-Zimmermann mass terms  in order to discuss  the gauge independence of the physical quantity.

\section{Nielsen Identities} 
To study the Nielsen identities we extend the action by introducing
  a global Grassmannian variable,  $\chi$ as follows
\begin{eqnarray}
\Sigma_{eff}^{m\prime}= \Sigma_{eff}^{m}+\int d^3x\ \frac{\chi}{2} \bar c b.\label{ac}
\end{eqnarray}
Here we remark that 
  the addition of this term does not change the dynamics
of the theory because of the Grassmannian nature of  $\chi$.
This extended action  (\ref{ac}) remains unchanged under the following set of  extended 
BRST transformations:
\begin{eqnarray}
\delta^+_b A_\mu &=&\frac{1}{e} \partial_\mu c\ \epsilon, \ \ 
\delta^+_b \psi  = ic\psi\ \epsilon, \nonumber\\  
\delta^+_b \bar\psi  &=& -ic\bar\psi\ \epsilon, \ \  
\delta^+_b c  = 0, \nonumber\\  
\delta^+_b \bar c  &=& \frac{1}{e} b\  \epsilon,  \ \  
\delta^+_b b  =0\nonumber\\
\delta^+_b \xi &=& -\frac{1}{e}\chi\ \epsilon,\ \ \delta^+_b \chi =0,
\end{eqnarray}
where $\epsilon$ is a transformation parameter.
Now, to derive the Nielsen identities we define the
following path integral with sources
\begin{eqnarray}
Z=\int {\cal D}\phi \exp{i\left[\Sigma_{eff}^{m} +\int d^3x \left(J_\mu A^\mu +\bar J_\psi \psi +\bar \psi J_{\bar \psi} +bJ_b +\bar \Omega (-ic\psi) +ic\bar \psi \Omega  \right)   \right]},
\end{eqnarray}
where various $J$ are the sources respective to associated fields and composite fields.
Now we define the vertex functional as follows
\begin{eqnarray}
\Delta (A_\mu, \psi, \bar{\psi}, b, c, \bar c, \chi, \xi, \bar \Omega, \Omega) =
W(J_\mu, J_{\bar{\psi}}, \bar J_\psi, J_b,\Omega,\bar{\Omega}, \chi, \xi) -\int d^3x(J_\mu A^\mu +\bar J_\psi \psi +\bar \psi J_{\bar \psi} +bJ_b),
\end{eqnarray}
where the generating functional $W$ generates only connected Green's function.
To study the gauge dependence of the  propagators, we now
introduce the functional integral of proper Green functions
\begin{eqnarray}
\delta_b^+ A_\mu \frac{\delta \Delta}{\delta A_\mu}
+\delta_b^+ \psi \frac{\delta \Delta}{\delta \psi}+\delta_b^+ \bar\psi \frac{\delta \Delta}{\delta \bar\psi}+\delta_b^+ \bar c \frac{\delta \Delta}{\delta \bar c}
+\delta_b^+ \xi \frac{\delta \Delta}{\delta \xi} =0.
\end{eqnarray}
Now we demand the invariance of the above functional under the extended BRST transformations
which yields 
\begin{eqnarray}
\frac{1}{e}\partial_\mu c \frac{\delta \Delta}{\delta A_\mu}
+\frac{\delta \Delta}{\delta \bar \Omega} \frac{\delta \Delta}{\delta \psi}+\frac{\delta\Delta}{\delta \Omega} \frac{\delta \Delta}{\delta \bar\psi}+ \frac{1}{e}b \frac{\delta \Delta}{\delta \bar c}
-\chi \frac{\delta \Delta}{\delta \xi} =0.
\end{eqnarray}
Differentiation  of above equation
  with respect to $\chi$ and then set $\chi$ equal to zero results
\begin{eqnarray}
\frac{\delta \Delta}{\delta\xi} +\frac{1}{e} \partial_\mu c \frac{\delta^2\Delta}{\delta\chi\delta A_\mu}-\frac{\delta^2\Delta}{\delta\chi\delta \bar\Omega}\frac{\delta\Delta}{\delta\psi}+\frac{\delta\Delta}{\delta\bar \Omega}\frac{\delta^2\Delta}{\delta\chi\delta\psi}
-\frac{\delta^2\Delta}{\delta\chi\delta \Omega}\frac{\delta\Delta}{\delta\bar\psi}+\frac{\delta\Delta}{\delta\Omega}\frac{\delta^2\Delta}{\delta\chi\delta\bar \psi}
-\frac{1}{e}b\frac{\delta^2\Delta}{\delta\chi\delta\bar c}=0.
\end{eqnarray}
This is the most general expression for the   Nielsen identities for the  QED$_3$ with the Lowenstein-Zimmermann mass terms.
From these expression we can generate the   Nielsen identities for the two-point functions.
With this result  one can  show  that   the
pole mass of the electron is gauge independent and that
the photon self-energy can be simply shown to be gauge
parameter independent.
  \section{FFBRST transformation} 
   
Let us review the FFBRST formulation \cite{bm}  in brief.
To do so,  we first write the 
usual BRST transformation for a generic field $\phi$ written collectively for massless
QED$_3$ theory,
\begin{equation}
\delta_b  \phi =s_b\phi\ \epsilon ={\cal R}[\phi] \epsilon,
\end{equation}
  where ${\cal R}[\phi]=s_b \phi$ is Slavnov variation of $\phi$
  and $\epsilon$ is infinitesimal parameter of transformation. 
  The importance of the  BRST transformation  does not alter by
considering  (i) the finite or infinitesimal and (ii) the field-dependent or field-independent versions of the  parameter $\delta\Lambda$  provided 
  the parameter must be anticommuting and space-time independent. This observation  gives us a freedom to 
generalize the BRST transformation by making the parameter, $\epsilon$, finite and field-dependent.  We first define the  infinitesimal field-dependent transformation as follows \cite{bm}
\begin{equation}
\frac{d\phi(x,\kappa)}{d\kappa}={\cal R} [\phi (x,\kappa ) ]\Theta^\prime [\phi (x,\kappa ) ],
\label{diff}
\end{equation}
where the $\Theta^\prime [\phi (x,\kappa ) ]$ is an infinitesimal  field-dependent parameter.
The FFBRST transformation ($\delta_f$) with the finite field-dependent parameter then can be 
obtained by integrating the above transformation from $\kappa =0$ to $\kappa= 1$, as follows:
 \begin{equation}
\delta_f \phi(x)\equiv \phi (x,\kappa =1)-\phi(x,\kappa=0)={\cal R}[\phi(x) ]\Theta[\phi(x) ],
\end{equation}
where 
$\Theta[\phi(x)]$
 is the finite field-dependent parameter constructed from its infinitesimal version. 
Under such FFBRST transformation with finite field-dependent
 parameter  the measure of generating function will  not be invariant
 and will contribute some non-trivial terms to the generating function in general \cite{bm}.

The Jacobian of the path integral measure $({\cal D}\phi)$ in the functional 
integral for such transformations is then evaluated for some 
particular choices of the finite field-dependent parameter, $\Theta[\phi(x)]$, as follows
\begin{eqnarray}
{\cal D}\phi^\prime &=&J(
\kappa) {\cal D}\phi(\kappa).\label{jacob}
\end{eqnarray}
Now, we  replace the Jacobian $J(\kappa )$ of the path integral measure  as   
\begin{equation}
J(\kappa )\longmapsto e^{{i\Sigma_1 [\phi(x,\kappa) ]}},\label{js}
\end{equation}
 iff the following condition \cite{bm}
 \begin{eqnarray}
 \int {\cal{D}}\phi (x) \;  \left [\frac{d}{d\kappa}\ln J(\kappa)-i\frac
{d\Sigma_1[\phi (x,\kappa )]}{d\kappa}\right ] \exp{[i(\Sigma_{eff}+\Sigma_1)]}=0 \label{mcond}
\end{eqnarray} 
is satisfied 
where $ \Sigma_1[\phi ]$ is some local functional of fields satisfying 
initial boundary condition   $ \Sigma_1[\phi ]|_{\kappa=0}=0$.
 
Moreover, the infinitesimal change in Jacobian, $J(\kappa)$,  is calculated by \cite{bm}
\begin{equation}
 \frac{d}{d\kappa}\ln J(\kappa)=-\int  d^3y\left [\pm\sum_i {\cal R}[\phi^i(y )]\frac{
\partial\Theta^\prime [\phi (y,\kappa )]}{\partial\phi^i (y,\kappa )}\right],\label{jac}
\end{equation}
where, for bosonic fields,  $+$ sign is used and for fermionic fields,
$-$ sign is used.
      
Therefore, by constructing an appropriate $\Theta$, we can calculate the non-trivial (local) Jacobian
 which extends the effective action by a term $\Sigma_1$.

\section{ Emergence of Lowenstein-Zimmermann mass terms}

In this section, we explicitly  show the emergence of Lowenstein-Zimmermann mass terms
for the massless QED$_3$ theory under FFBRST formulation.
To implement these notions, we construct the FFBRST transformation corresponding to Eq.  (\ref{BRS}) 
following the techniques outlined in Sec. III:
\begin{eqnarray}
\delta_f A_\mu &=&\frac{1}{e} \partial_\mu c\ \Theta[\phi], \nonumber\\ 
\delta_f \psi  &=& ic\psi\ \Theta[\phi], \nonumber\\  
\delta_f \bar\psi  &=& -ic\bar\psi\ \Theta[\phi], \nonumber\\  
\delta_f c  &=& 0, \nonumber\\  
\delta_f \bar c  &=& \frac{1}{e} b\ \Theta[\phi], \nonumber\\  
\delta_f b  &=& 0, 
\end{eqnarray}
where $ \Theta[\phi]$ is an arbitrary field-dependent parameter of transformation.
It is easy to check that the effective action for QED$_3$ given in (\ref{eff}) is invariant 
under these set of  transformations. Now, we  construct
a particular field-dependent parameter (following the procedure given in Ref. \cite{bm}) as
\begin{equation}
\Theta[\phi] =  \int d^3x\ e\frac{\bar c b  }{b^2} \left[\exp\left(i\frac{\mu}{2}(s-1)\epsilon^{\mu\nu\rho}A_\mu \partial_\nu A_\rho +i m(s-1)\bar\psi \psi +i\psi\bar\Omega  -i\bar{\psi} \Omega 
\right)-1\right]. \label{the}
\end{equation}
Now, following the method discussed in section III, we calculate the  Jacobian for path integral measure 
under finite field-dependent 
BRST transformation with above parameter as follows 
 \begin{equation}
J[\phi (x)] =e^{i\Sigma_1}= e^{i\int d^3x \left[ \frac{\mu}{2}(s-1)\epsilon^{\mu\nu\rho}A_\mu \partial_\nu A_\rho + m(s-1)\bar\psi \psi + \bar{\Omega}s_b\psi - s_b\bar{\psi}\Omega\right]}.  
\end{equation}
Here to obtain this expression we have utilized the relation (\ref{mcond}).

As a consequence of performing FFBRST transformation on path integral measure of 
functional integral we see that the effective action of the theory gets extendend (within functional integral) as follows:
\begin{eqnarray}
\Sigma_{eff} + \Sigma_1 &=& \int  d^3x\left[-\frac{1}{4} F_{\mu\nu}F^{\mu\nu} + i\bar\psi  D \psi + b\partial^\mu A_\mu + \frac{\xi}{2}b^2 + \bar c \partial_\mu \partial^\mu c \right.\nonumber\\
&+&\left.\frac{\mu}{2}(s-1)\epsilon^{\mu\nu\rho}A_\mu \partial_\nu A_\rho + m(s-1)\bar\psi \psi + \bar{\Omega}s_b\psi - s_b\bar{\psi}\Omega \right].
\end{eqnarray}
 which   exactly coincides with effective action given in (\ref{eff}).
This justifies our claim of emergence of Lowenstein-Zimmermann mass terms
naturally under the celebrated FFBRST technique. We also emphasized that
the external source terms for the non-linear BRST variations which are required 
to prove the renormalizability of the theory are automatically generated through
the same FFBRST transformation.
  
It means that under FFBRST transformation with appropriately constructed parameter (\ref{the})
the effective action (within functional integral) changes as
\begin{eqnarray}
\Sigma_{eff}+\Sigma_1 =\Sigma_T.
\end{eqnarray}
Hence, the whole mechanism is, precisely, given by
\begin{eqnarray}
\int {\cal D\phi}\ e^{i\Sigma_{eff}}\stackrel{FFBRST}{----\longrightarrow}
\int {\cal D\phi}\ e^{i\left(\Sigma_{eff}+\Sigma_1\right)}=
\int {\cal D\phi}\ e^{i\left(\Sigma_{eff}+\Sigma_m+\Sigma_{ext}\right)},
\end{eqnarray}
where the generic path integral measure (${\cal D\phi}$) is explicitly given by
${\cal D\phi}= {\cal DA_\mu}{{\cal D}b}{{\cal D}c}{{\cal D} \bar{c}}{{\cal D}\psi}{{\cal D}\bar\psi}$. 
Therefore,  the  Lowenstein-Zimmermann mass terms and the external source 
term are generated through the Jacobian of the path integral measure under generalized BRST
transformations
with appropriate transformation parameter.
\section{Conclusions}
 In this paper we have studied the BRST symmetry for the ultraviolet finite, super-renormalizable theory of 
 massless QED$_3$. Also we have derived the Nielsen identities for   QED$_3$.
  The
Nielsen identities are important to investigate because these offer possibilities to check one's calculations as they  allow us to see where physical
meaning may be found in apparently gauge dependent Green's functions.  
 Further we have generalized the BRST symmetry of the theory by making the transformation parameter finite 
 and
 field dependent which is known as FFBRST transformation.  
 The  fascinating feature of FFBRST transformation is that
 under   change of variables it leads to a non-trivial Jacobian for the 
 path integral measure of generating functional. This Jacobian, 
  written as $e^{iS_1}$ for some local functional
 of fields  $S_1$,    depend on the choice of finite 
 field-dependent parameter. We have computed the Jacobian for FFBRST transformation 
 with appropriate finite field-dependent parameter. 
  Remarkably, we have found that the   Lowenstein-Zimmermann mass terms together with the external sources 
 for massless QED$_3$  emerges naturally  within functional integral
 through the Jacobian of a single FFBRST transformation.
 This is remarkable feature of FFBRST symmetry that any  gauge   invariant (BRST exact)  quantity
can be generated through the FFBRST symmetry. Although  these  Lowenstein-Zimmermann  terms
are  mass terms but   are gauge invariant also. 
Thus we have seen that the extra physical degree's of freedom emerges due to
the non-linear BRST transformations (23) where the parameter $\Theta$ exhibitis  
the extra physical degrees of freedom due to the mass terms. Though we illustrated 
our results for QED$_3$ theory but certainly these are not limited to a particular theory. However it is a more general result 
and can be applied to any gauge theory to get gauge invariant mass terms and there dynamics.


\begin{thebibliography}{99}
\bibitem{del} O. M. Del Cima,  D. H. T. Franco   and O. Piguet, Phys. Rev. D 89, 065001 (2014).
\bibitem{gua} E. Guadagnini, M. Martellini and M. Mintchev, Phys. Lett. B 227, 111 (1989).
\bibitem{gua1} E. Guadagnini, M. Martellini and M. Mintchev, Nucl. Phys. B 330, 575 (1990).
\bibitem{gua2} A. Blasi and R. Collina, Nucl. Phys. B 345, 472 (1990).
\bibitem{gua3} N. Maggiore and S. P. Sorella, Nucl. Phys. B 377, 236 (1992).
\bibitem{nd} N. Dorey and N.E. Mavromatos, Nucl. Phys. B 386, 614 (1992).
\bibitem{kf} K. Farakos and N.E. Mavromatos, Mod. Phys. Lett. A 13, 1019 (1998).
\bibitem{pen} M.R. Pennington and D. Walsh, Phys. Lett. B 253, 246 (1991).
\bibitem{lee} D. Lee  and P. Maris, Phys. Rev. D 67, 076002 (2003).
\bibitem{jh} J. H. Lowenstein, Phys. Rev. D 4, 2281 (1971).
\bibitem{jh1} J. H. Lowenstein,  Commun. Math. Phys. 24, 1 (1971).
 \bibitem{jh2} Y.M. P. Lam, Phys. Rev. D 6, 2145 (1972).
 \bibitem{jh3}  Y.M. P. Lam, Phys. Rev. D   7, 2943 (1973).
 \bibitem{jh4}   T. E. Clark and J. H. Lowenstein, Nucl. Phys. B113, 109 (1976).
\bibitem{brs} C. Becchi, A. Rouet, and R. Stora, Comm. Math. Phys. 42,
127 (1975).
\bibitem{brs1} C. Becchi, A. Rouet, and R. Stora,  Ann. Phys. (N.Y.) 98, 287 (1976);
 \bibitem{brs2} O. Piguet
and A. Rouet, Phys. Rep. 76, 1 (1981).
\bibitem{jh5}J. H. Lowenstein and W. Zimmermann, Nucl.
Phys. B86, 77 (1975); J.H. Lowenstein, Comm. Math.
Phys. 47, 53 (1976).
\bibitem{nil} N. K. Nielsen, Nucl. Phys. B 101, 173 (1975).
\bibitem{del1} R. Fukuda and T. Kugo, Phys. Rev. D 13, 3469 (1976).
 \bibitem{del2}I. J. R. Aitchison and C. M. Fraser, Ann. Phys. 156, 1 (1984).
\bibitem{del3}  O. M. Del Cima, D. H. T. Franco and O. Piguet, Nucl.Phys. B 551, 813 (1999).
\bibitem{masud}M. Chaichian and  N. F. Nelipa,  \textit{Introduction to Gauge Field Theories},
Springer Berlin Heidelberg, (2012).
\bibitem{bm} S. D. Joglekar and B. P. Mandal, Phys. Rev. D 51, 1919 (1995).
\bibitem{mru} B. P. Mandal, S. K. Rai and S. Upadhyay, EPL 92, 21001 (2010).
\bibitem{sdj1}  S. D. Joglekar and B. P. Mandal, Int. J. Mod. Phys. A 17, 1279 (2002).
\bibitem{rb} R. Banerjee and B. P. Mandal, Phys. Lett. B 488, 27 (2000).
 \bibitem{susk}   S. Upadhyay,   S. K. Rai and B. P. Mandal,  J. Math. Phys.  {52}, {022301} (2011).
 \bibitem{jog} S. D. Joglekar and A. Misra, Int. J. Mod. Phys. A 15, 1453 (2000).

\bibitem{sb1} S. Upadhyay and B. P. Mandal,  Eur. Phys. J.  {C 72},  2065 (2012).
\bibitem{sb2}  S. Upadhyay and B. P. Mandal,  Annals of Physics {327}, 2885 (2012).
\bibitem{sb3}   S. Upadhyay and B. P. Mandal,     Mod. Phys. Lett.   {A 25}, {3347} (2010).
\bibitem{sb4}   S. Upadhyay and B. P. Mandal,     AIP Conf. Proc. 1444, 213 (2012).
 \bibitem{sb5}  S. Upadhyay and B. P. Mandal,      Prog. Theor. Exp. Phys.   053B04,  (2014).
  \bibitem{sb6} S. Upadhyay and B. P. Mandal,       Eur. Phys. J. C 75,  327 (2015).
 \bibitem{sb7}   S. Upadhyay and B. P. Mandal,        arXiv:1503.07390.
  \bibitem{sb8}   S. Upadhyay and B. P. Mandal,        Phys. Lett. B 744, 231 (2015). 

\bibitem{smm} S. Upadhyay, M. K. Dwivedi and B. P. Mandal, Int. J. Mod. Phys. A 28, 1350033 (2013).
\bibitem{sudd} S. Upadhyay and D. Das, Phys. Lett. B 733, 63 (2014).
\bibitem{fsm}M. Faizal, S. Upadhyay and B. P. Mandal, Phys. Lett. B 738, 201 (2014).
\bibitem{fsm1}M. Faizal, S. Upadhyay and B. P. Mandal,  Int. J. Mod. Phys. A 30, 1550032 (2015).
\bibitem{sud} S. Upadhyay, Phys. Lett. B 727, 293 (2013).
\bibitem{sud1} S. Upadhyay, EPL  104, 61001  (2013).
\bibitem{sud2} S. Upadhyay,  EPL 105, 21001 (2014).
\bibitem{sud3} S. Upadhyay, Annals of Physics 344, 290 (2014).
\bibitem{sud4} S. Upadhyay, Annals of Physics 340, 110 (2014).
\bibitem{sud5} S. Upadhyay,  Phys. Lett. B 740, 341   (2015).
 \bibitem{sud6} S. Upadhyay,  Mod. Phys. Lett. A 30, 1550072 (2015).
 \bibitem{sud7} S. Upadhyay,   Annals of Physics 356, 299 (2015).
\bibitem{fs} M. Faizal, B. P. Mandal and S. Upadhyay, Phys. Lett. B 721, 159 (2013).
\bibitem{rbs} R. Banerjee, B. Paul and S. Upadhyay,  Phys. Rev. D 88, 065019 (2013).
\bibitem{rbs1} R. Banerjee  and S. Upadhyay, Phys. Lett. B 734, 369 (2014).
 \bibitem{gri} V. N. Gribov, Nucl. Phys. B 139, 1 (1978).
\bibitem{zwan}  D.Zwanziger, Nucl. Phys. B 323, 513 (1989).
 \bibitem{sb} S. Upadhyay and B. P. Mandal, EPL 93, 31001 (2011).
\bibitem{sb20} S. Upadhyay and B. P. Mandal, AIP Conf. Proc. 1444, 213 (2012).
\bibitem{lav} P. M. Lavrov and O. Lechtenfeld,  Phys. Lett. B 725,  382 (2013).
\bibitem{ale} P. Y. Moshin and A. A. Reshetnyak,  arXiv: 1405.0790.
\bibitem{ale1} P. Y. Moshin and A. A. Reshetnyak,  arXiv:  1406.5086.
\bibitem{ale2} P. Y. Moshin and A. A. Reshetnyak,  arXiv: 1405.7549.
\bibitem{bata} I. A. Batalin, P. M. Lavrov and I. V. Tyutin,   arXiv: 1404.4154. 
\bibitem{bata1} I. A. Batalin, P. M. Lavrov and I. V. Tyutin, arXiv: 1405.7218.
\end{thebibliography}
\end{document}